\documentclass[amsmath,twocolumn, amssymb,aps,floatfix,nolongbibliography]{revtex4-2}

\usepackage{graphicx,overpic}
\usepackage{dcolumn}
\usepackage{bm}

\begin{document}
	\newcommand{\RNum}[1]{\uppercase\expandafter{\romannumeral #1\relax}}

\title{Hierarchical Topological States in Thermal Diffusive Networks}
\author{Biao Chen$^{1,2}$}
\author{Kaiyun Pang$^{1}$}
\author{Ru Zheng$^{1}$}
\email{Zhengru@nbu.edu.cn}
\author{Feng Liu$^{1,2,3}$}
\email{Liufeng@nbu.edu.cn}

\affiliation{$^{1}$School of Physical Science and Technology, Ningbo
  University, Ningbo, 315-211, China}
 
\affiliation{$^{2}$Institute of High Pressure Physics, Ningbo
  University, Ningbo, 315-211, China}

\affiliation{$^{3}$Department of Nanotechnology for Sustainable Energy, School of Science and Technology, Kwansei Gakuin University, Gakuen 2-1, Sanda 669-1337, Japan}

\begin{abstract}
The integration of topological concepts into electronic energy band theory has been a transformative development in condensed matter physics. 
Since then, this paradigm has broadened its reach, extending to a variety of physical systems, including open ones. 
In this study, we employ analogues of the generalized $n$-dimensional Su-Schrieffer-Heeger model, a cornerstone in understanding topological insulators and higher-order topological states, to unveil a dimensional hierarchy of topological states within thermal diffusive networks. 
Unlike their electronic counterparts, the topological states in these networks are characterized by confined temperature profiles of dimension $(n-d)$ with constant diffusive rates, where $n$ represents the system's dimension and $d$ is the order of the topological state. 
Our findings demonstrate the existence of topological corner states in thermal diffusive systems up to $n=3$, along with surface and hinge states. 
We also identify and discuss an intermediate-order topological phase in the case $n=3$, characterized by the presence of hinge states but the absence of corner states. Furthermore, our work delves into the influence of chiral symmetry in these thermal networks, particularly focusing on topological thermal states with a near-zero diffusion rate. 
This research lays the foundation for advanced thermal management strategies that utilize topological states in multiple dimensions.

\end{abstract}

\maketitle

\section{Introduction}
The field of condensed matter physics has been revolutionized by integrating topological concepts into the theory of electronic energy bands~\cite{Hasan2010, Qi2011, Ando2013, Bansil2016, Rachel2018}. This transformation began with the pioneering discovery of topological insulators, which fundamentally altered our understanding of electronic systems and catalyzed extensive research into topological phenomena across diverse physical domains, including photonic~\cite{Fangkejie2012, Lu2014, Chen2014, Khanikaev2017, Ozawa2019, Chenjianfeng2022}, acoustic~\cite{He2016, Xia2018, Ma2019, Peri2020}, and mechanical metastructures~\cite{Roman2015, Kariyado2015, Xue2022}, as well as electrical circuits~\cite{Jia2015, Luo2018, Lee2018, Liushuo2019, Wangzhu2023}. These topological parallels across different physical platforms are not merely academic interests; they pave the way for practical applications. By enabling control and manipulation of classical waves and currents in a manner similar to electronic topological materials, they hold great promise for applications like topological quantum computing and efficient, scatter-free wave transport.

The fundamental concept in energy band topology is the bulk-edge correspondence~\cite{Hatsugai1993, Liang2006, Hwang2019, Bouhon2019, Wang2021, Yoshida2021}. This principle predicts the presence of robust states at the interfaces between topologically distinct domains. 
Recent advances have expanded this concept into a higher-order framework, establishing a $n-(n-d)$ correspondence, where $n$ is the dimension of the topological bulk system and $d$ the order of the topological states, as confirmed through both theory and experimentation~\cite{Liu2017, Benalcazar2017a, Benalcazar2017B, Song2017, Langbehn2017, Khalaf2018, Geier2018, Ezawa2018, Schindler2018, Schindler2018a, Xie2018, Ezawa2018a, Wang2018a, Zhu2018, Queiroz2019, Yan2019, Zhang2019b, Luo2019, Xie2019, Liu2019, Park2019, Yue2019, Okugawa2019, LiuBing2019, Liu2019a, Zhang2019d, Trifunovic2019, Peng2020, Cerjan2020, Chen2020, Hua2020, Hu2020, Huang2020, Choi2020, Nag2021, Aggarwal2021, Ghosh2021, Khalaf2021, Wang2021a, Zhang2021a, Liu2021a, Benalcazar2022, Tan2022, Lei2022, Jia2022}.  
Higher-order topological materials differ from conventional ones by exhibiting nontrivial states not just at boundaries, but also at 'boundaries of boundaries', like hinges and corners.

The advent of these higher-order states opens new avenues for designing robust and efficient devices for various applications, including nanocavity-based sensors and topological lasers~\cite{Hararieaar2018, Benalcazar2019, Wu2020, Watanabe2021, Jung2021, Zhang2020c, Zhang2020d, Pahomi2020, Pan2022, Tang2023, He2023}. Illustrative examples, such as the two-dimensional Su-Schriffer-Heeger (SSH) model~\cite{Liu2017}, the Benalcazar–Bernevig–Hughes model~\cite{Benalcazar2017a, Benalcazar2017B}, and the breathing Kagome lattice~\cite{Ezawa2018}, demonstrate the realization of higher-order topological states in diverse classical wave systems, including photonic crystals~\cite{Xie2019, Chen2019, Mittal2019}, acoustic resonators~\cite{Serra-Garcia2018, Xue2019, Lin2020}, and electronic circuits~\cite{Imhof2018,Liushuo2020}.

Very recently, topological edge states have been experimentally observed in 1D and 2D thermal diffusive lattices. Thermal diffusive topological lattices are realized on the basis of analogs of the 1D and 2D SSH models~\cite{Qi2022,Hu2022,Wu2023,Xu2023}. By tuning the heat diffusion rate between thermal "sites", confined temperature profiles emerge at the interface of two topologically distinct domains characterized by the (vectored) Zak phase~\cite{Liu2017, Liu2019}. It is important to note that, in contrast to the wave systems previously mentioned, a thermal diffusive process is intrinsically an open system. Moreover, the Hamiltonian analogue that governs the temperature field in these lattices is anti-Hermitian~\cite{Liying2019, Zhang2023}, leading to eigenspectra that are purely imaginary and correspond to the rate of thermal diffusion.


While 1D and 2D topological thermal lattices have been successfully implemented in experiments, the exploration of 3D thermal lattices is less advanced, possibly due to the challenges of constructing 3D thermal diffusive networks experimentally~\cite{Liuzhoufei2022}.
However, theoretical exploration of 3D topological thermal lattices reveals several critical areas for further research.
First, we need to confirm the hierarchical dimensional structure of higher-order topological states in these lattices.
Second, it is important to investigate the newly proposed intermediate-order topological phases, a unique phenomenon of recently proposed hierarchical topological insulators that manifest only in dimensions greater than two~\cite{Liu2023}.
Lastly, the complex nature of higher-dimensional structures provides increased flexibility, which offers a potentially greater scope for practical application in this field.

In this work, using analogs of the generalized $n$-dimensional ($n$D) SSH model, we study topological states, including higher-order ones in thermal networks up to $n=3$. 
We verify the dimensional hierarchy of higher-order topological states, that is, topological states of dimension $(n-d)$ appear consequently for $d=1$, $2$, $\cdots$, $n-1$. 
Particularly in the 3D case, we explore an intermediate-order topological phase characterized by the presence of surface and hinge states but a notable absence of corner states. 
These intermediate-order topological states are robust against perturbation along a specific direction with large amplitudes. 
Additionally, on the basis of the analog of the $n$-dimensional SSH model, we delve into the concept of chiral symmetry in diffusive systems, focusing on a thermally still topological state with a decaying rate almost zero in the 1D lattice.    


The remaining parts of the paper are organized as follows: 
In Sec.~\RNum{2}, we develop the tight-binding scheme for the thermal network based on Fourier's law. 
Section~\RNum{3} is dedicated to exploring the analogues of the
$n$D SSH model within thermal networks, where we analyze their topological properties for $n=1,2$ and 3. 
In Sec.~\RNum{4}, we examine the robustness of topological states in these thermal networks. 
The significance of chiral symmetry in relation to the zero decay rate in thermal networks is discussed in Sec.~\RNum{5}. Finally, the summary is given in Sec.~\RNum{6}. 

\section{Fourier's Law and Tight-binding Scheme of thermal network}
In a normal thermal diffusion process, Fourier's law governs the dynamics, which is written as
\begin{equation}
	\mathbf{J}=-k \nabla T,
\end{equation}
where $\mathbf{J}$ is the local heat flux density, $k$ is the material's thermal conductivity, $T$ is the temperature field, and $\nabla T$ is temperature gradient. 
Considering the first law of thermodynamics, or the energy conversation law, which is given by $c\rho\partial_t T=-\nabla \cdot \mathbf{J}$, Fourier's law can be reformulated into the heat equation:

\begin{equation}
\partial_t T=D \nabla^2 T,
\end{equation}
where $D=k/c\rho$ with $c$ the specific heat capacity and $\rho$ the medium density. 
Additionally, an external heat source $Q$ can be incorporated into the right-hand side of this equation. 
For instance, a convection term $h(T_\text{env}-T)/c\rho$ can be added, where $h$ and $T_\text{env}$ are the convection power and the environmental temperature.  

In our study, we conceptualize a thermal network where the temperature is uniformly distributed across each "site", connected by thermal conductive links. 
This is illustrated in Fig. ~1(a), where the circles represent the sites and the lines denote the thermal conductive links. 
For such a thermal network, $\nabla^2$ in the heat equation becomes a form of finite difference, tailored to each individual site. 
Furthermore, the diffusion coefficient $D$ varies from site to site, reflecting the topological essence of thermal networks. 
Adapting the heat equation for a thermal network, we have
\begin{equation}
\partial_t \mathbf{T}=
\begin{pmatrix}
    -\sum_jD^{1j}& D^{12} &\cdots& D^{1n}\\
    D^{21} &-\sum_jD^{2j}  &\cdots & D^{2n}\\
    \vdots& \ddots\\
    D^{n1} & D^{n2} & \cdots&-\sum_j{D^{nj}}
\end{pmatrix}
\begin{pmatrix}
T_1\\
T_2\\
\vdots\\
T_n
\end{pmatrix}
,
\label{TBM}
\end{equation}
where $\mathbf{T}=(T_1,T_2,\cdots,T_n)$ is the temperature field of each site, $D^{ij}$ is the thermal conductance from site $i$ to site $j$, and the summation is over all the connecting site $j$ of site $i$. If a convection term is taken into account, a constant term $-h/\rho c$ is added to the diagonal terms of Eq.~(\ref{TBM}), and the $T_\text{env}$ defines the base temperature of each site.

Assuming an exponentially decay of the temperature vector $\mathbf{T}$ over time as $e^{-\gamma t}$, we can reformulate Eq.~(\ref{TBM}) as 
\begin{equation}
\omega \mathbf{T}=-i \mathbf{D}\mathbf{T}, 
\end{equation}
where $\omega=i\gamma$ represents the decay rate in complex form, and $\mathbf{D}$ is the thermal linkage matrix. 
Intriguingly, Eq.~(\ref{TBM}) mirrors the form of the eigenvalue equation in quantum mechanics, suggesting an analogous Hamiltonian $\mathcal{H}=-i\mathbf{D}$. In the case of purely real $\mathbf{D}$, $\mathcal{H}$ is anti-Hermitian, owing to the incorporation of the imaginary prefactor. This analogy opens up a fascinating parallel between the behavior of thermal networks and quantum systems.

Equation (4) implies that, apart from the imaginary prefactor, the thermal network can be analogously treated as a tight-binding model used in solid-state materials. 
This approach enables us to transpose the concept of topological phases, common in energy band theory, to thermal diffusive networks. 
In subsequent sections, we demonstrate this by emulating the generalized $n$-dimensional SSH model, up to $n=3$, using thermal networks and examining the topological characteristics of thermal diffusion. 

For the numerical simulation of these thermal networks, we employ the finite element method using COMSOL. 
The material parameters selected for the simulations are representative of the aluminum alloy, with the heat capacity $c=978[\text{J}/(\text{kg}\cdot K)]$ , the mass density $\rho=2700[\text{kg/}m^3]$, and the thermal conductivity $k=180[W/\text{m}\cdot\text{K}]$. 
For all sections except Sec.~\RNum{5}, we assume an air circulation power of $18[\text{W}/\text{m}^2\cdot \text{K}]$.

\section{thermal networks of $n$D SSH Model}
\subsection{1D SSH thermal network}
\begin{figure}[!htbp]
	\centering
    \includegraphics[width=1\linewidth]{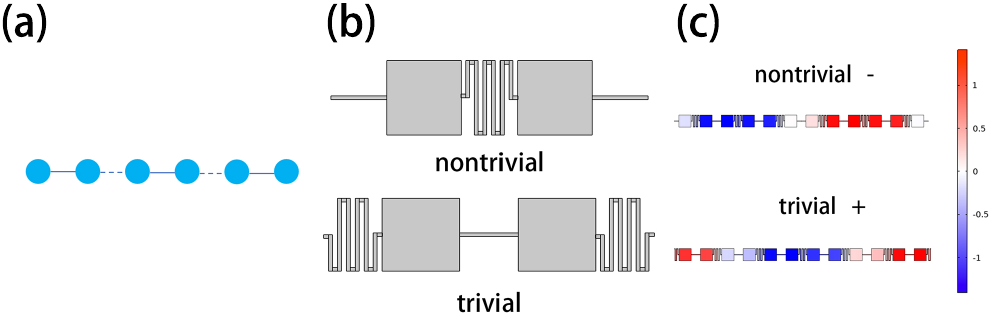}
\caption{(a) Schematic of thermal network mimicking 1D SSH model, where the solid and dashed lines indicate intra- and inter-cells thermal conductance. 
(b) Unit cells of 1D SSH thermal network for topological nontrivial and trivial phases. 
(c) Temperature profiles of 1D SSH thermal network within the first band for the nontrivial and trivial unit-cell under periodic boundary conditions.}
\end{figure}

The 1D SSH model, developed in the 1970s, is instrumental in explaining the electronic properties of polyacetylene, a polymer made of repeating ethylene units. 
A key feature of the 1D SSH model is its ability to exhibit topological properties in a relatively simple system, which is characterized by alternating hopping amplitudes: intracell hopping $\gamma$ and intercell hopping $\gamma^\prime$. 
The model includes two sites per unit cell, and its Hamiltonian is given by
\begin{equation}
    \mathcal{H}_{1D}=\sum_N( \gamma a_N^\dagger b_N + \gamma^\prime b^\dagger_{N-1}a_{N})+\text{h.c.},
    \label{1DSSH}
\end{equation}
where $N$ represents the index of unit-cell, and $a$, $b$ are two sites in each unit cell. 
Figure 1(a) provides a schematic representation of the 1D SSH model.

A topological phase transition in the 1D SSH model is marked by the condition $\gamma=\gamma^\prime$, leading to a band inversion at $k=\pi$. This inversion can be identified by the parity of the wavefunction, particularly when inversion symmetry is present, such as exhibiting odd parity for the lowest band. 
This diagnostic approach is applicable for determining the topological nature of thermal networks. 
As predicted by the bulk-edge correspondence, topological states arise at the interface between two domains of distinct topological characteristics.

Drawing from Eqs.~(\ref{TBM}) and (\ref{1DSSH}), we construct an analog thermal network of the 1D SSH model. 
In this network, the alternating hopping amplitudes of the SSH model are represented by distinct elements in the matrix $D$. 
The two types of unit cells of the 1D SSH thermal network are illustrated in Fig.~1(b). 
Here, the two distinct structures that connect sites correspond to the non-equivalent intracell hopping $D_x$ and intercell hopping $D^\prime_x$.

The topological characteristics of this network can be identified by examining the parity of the temperature profile in a finite chain under periodic boundary conditions. 
As displayed in Fig.~1(c), for the case $D_x<D_x^\prime$, the temperature profile of the lowest band has an odd parity, while for $D_x>D_x^\prime$, it has an even parity, consistent with the 1D SSH model of Eq.~(\ref{1DSSH}). 
In our model, each site has a length of 2 cm, and the distance between adjacent sites is also set to 2 cm. The thermal conductance between each site is determined by the total length and the cross-sectional area of the thermal link.

\begin{figure}[t]
	\centering
	\includegraphics[width=1\linewidth]{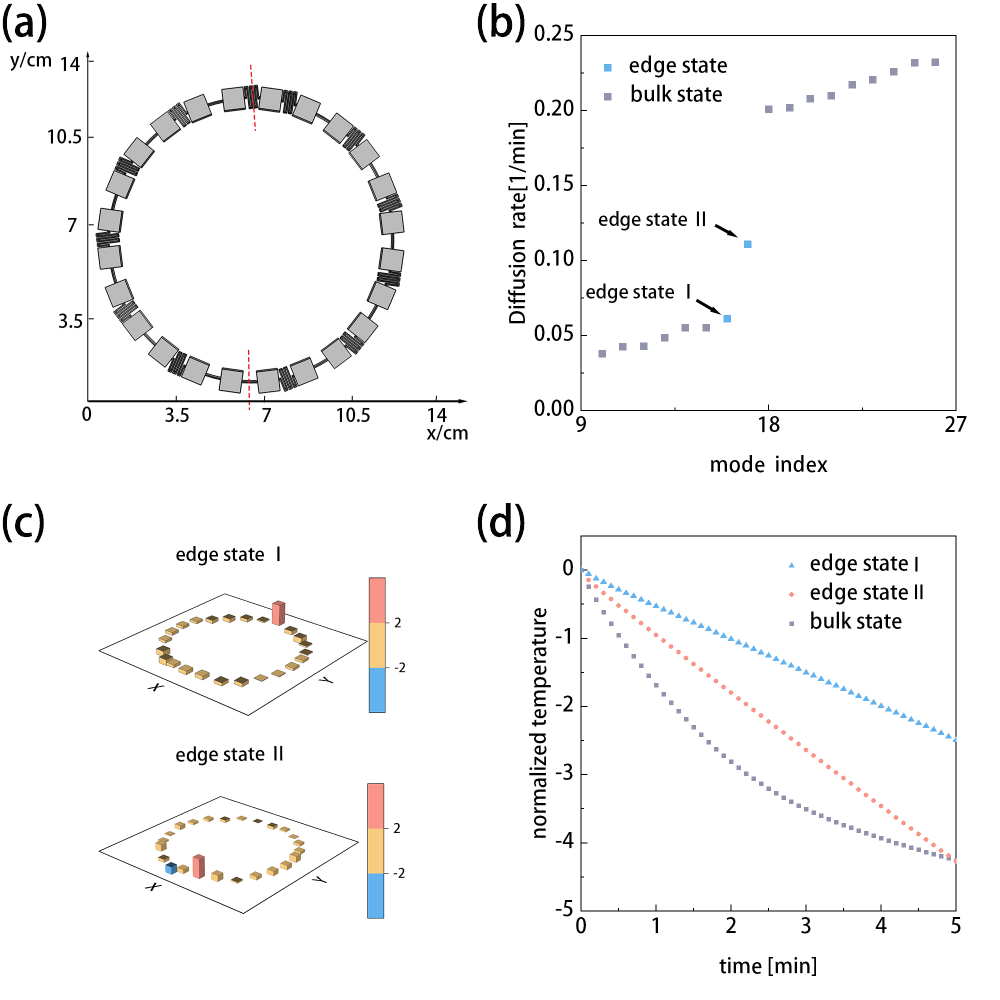}

\caption{(a) Thermal network structure made up of two chains of 1D SSH model with trivial and nontrivial unit cells. Dashed lines indicate the positions of topological interfaces. 
(b) Energy band spectrum of the thermal network in (a). Topological edge states appear in the band gap.
(c) Temperature profiles of two topological edge states.
(d) Time evolution of bulk and edge states in (b).}
\end{figure}

Having established the unit cell of the 1D SSH thermal network with distinct topological labels, we now turn our attention to examining its topological properties, including topological edge states.
To this end, we construct a ring structure, as depicted in Fig.~2(a), with dashed lines indicating the interfaces between two topologically different domains. By solving the eigenproblem using COMSOL for the ring structure shown in Fig.~2(a), we determine the characteristic diffusion rates of the thermal network.

Consistent with the behavior of the 1D SSH model, we observe the emergence of a band gap in the characteristic spectrum, as illustrated in Fig.~2(b).  This gap arises due to the disparity between $D_x$ and $D^\prime_x$. 
Notably, within the band gap, two edge states emerge. 
These edge states are located at each of the topological interfaces of the ring structure. 
The temperature profiles of these two topological edge states, shown in Fig.~2(c), reveal a localized temperature profile at the interface sites. It is important to note that the thermal linkage between the two distinct topological domains is adjustable. This adjustability allows us to tune the diffusion rate of the edge state within the band gap, providing insight into the dynamic behavior of topological edge states in thermal networks.

A key characteristic that distinguishes diffusive edge states from bulk states is their relative isolation. 
As a result, their decay rates exhibit remarkable stability over time. 
As displayed in Fig.~2(d), we see that after 5 minutes of time evolution, the decay rates of two edge states remain almost constant. 
This behavior contrasts with that of bulk diffusive states, which tend to become a mixture of several states over time. The normalized temperature is defined as $\ln(\frac{T-T_\text{env}}{T_0-T_\text{env}})$ with $T_0$ the initial temperature.

Such stability in the decay rates of edge states opens up new possibilities for controlling heat diffusion processes by utilizing topological states. Furthermore, when combined with chiral symmetry, we can even achieve a "still" thermal diffusive state, a phenomenon that we explore in greater detail in subsequent sections.

\subsection{2D SSH thermal network}

\begin{figure}[b]
	\centering
	\includegraphics[width=1\linewidth]{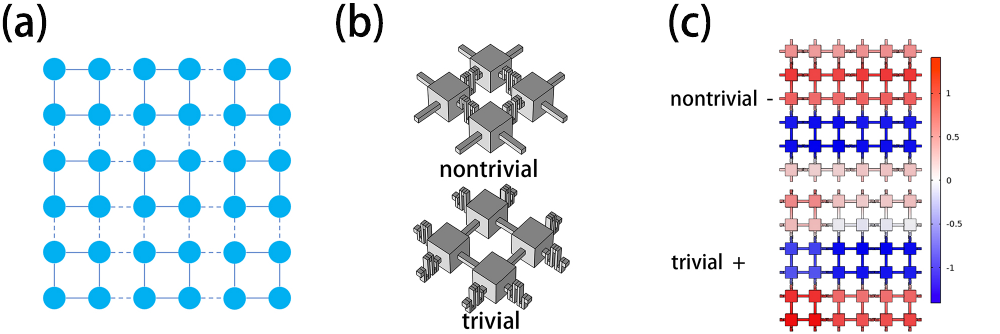}

\caption{(a) Schematic of thermal network mimicking 2D SSH model, where the solid and dashed lines indicate intra- and inter-cells thermal conductance. 
(b) Unit cells of 2D SSH thermal network for topological nontrivial and trivial phases. 
(c) Temperature profiles of 2D SSH thermal network within the first band for the nontrivial and trivial unit-cell under periodic boundary condition.}
\end{figure}

The 2D Su-Schriffer-Heeger (SSH) model introduces a novel concept in topological physics: the higher-order topological phase. Unlike conventional topological states, higher-order topological states are characterized by codimensions greater than one. A prime example is the corner states found in 2D systems.

Building on the 1D SSH model, the 2D SSH model incorporates hopping amplitudes that depend on two spatial dimensions.  
These include $\gamma_x$, $\gamma_y$, $\gamma_x^\prime$, and $\gamma_y^\prime$. 
Figure 3(a) provides a schematic representation of the 2D SSH model, illustrating its spatially dependent hopping amplitudes along the $x$- and $y$- directions.
Incorporating the alternating hopping amplitudes along the $x-$ and $y-$ directions, we construct the unit cells for the 2D SSH thermal network, as depicted in Fig.~3(b). 
In this setup, we focus on the symmetric case of $D_x=D_y=D$ and $D^\prime_x=D^\prime_y=D^\prime$.
To classify the topological nature of unit cells, we analyze the temperature profiles within the lowest band, shown in Fig.~3(c).  
By comparison of the amplitude between $D$ and $D^\prime$ and the parity of the temperature profiles, we can label the topological class for the two types of unit cells illustrated in Fig.~3(b). It should be noted that, due to the symmetric thermal conductance structures, we can focus on the parity for the $x$-axis only.   

To show the higher-order topological properties within the 2D SSH thermal network, we construct a finite 2D sample composed of two topologically different domains, as illustrated in Fig.~4(a).  
In this figure, the dashed square encloses a section of the 2D thermal network formed from the nontrivial unit cells of the 2D SSH model. 
In contrast, the area outside the square comprises the trivial unit cells. 
To minimize the influence of finite-size effects, we implement periodic boundary conditions along both the $x$- and $y$-axes.

\begin{figure}[t]
	\centering
	\includegraphics[width=1\linewidth]{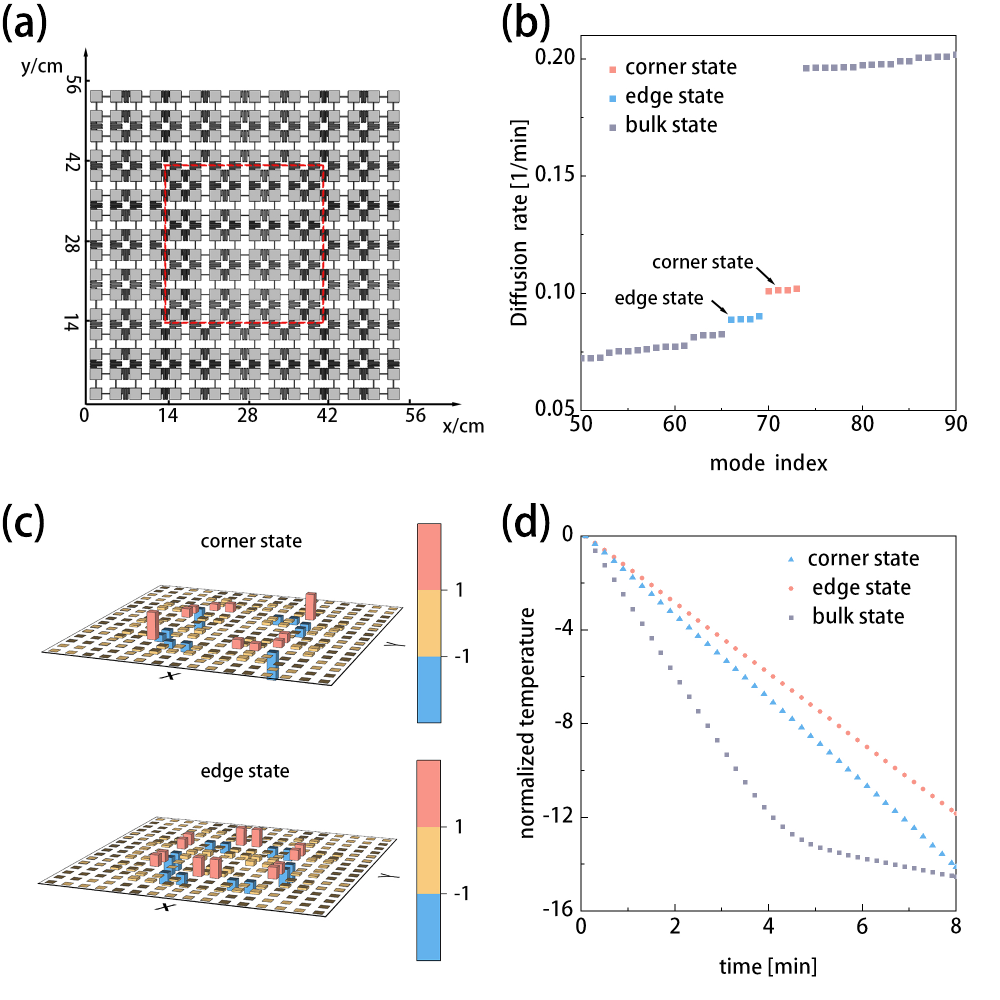}

\caption{(a) Thermal network structure made up of the 2D SSH models with trivial and nontrivial unit-cells, where the dashed lines indicate the positions of topological interfaces. 
(b) Energy band spectrum of the thermal network in (a). Topological edge and corner states appear in the band gap.
(c) Temperature profiles of the topological edge states.
(d) Time evolution of bulk, edge, and corner states in (b).}
\end{figure}

The characteristic decay rates of the 2D SSH model are shown in Fig.~4(b). As with the 1D SSH thermal network, a band gap is evident in the decay rate spectrum.
It is noted that because of the continuous nature of the thermal equation, the decay rate spectrum can go beyond the first band gap, and here we consider the first band gap only. 
At the topological interfaces marked by the dashed square in Fig.~4(a), topological states emerge within this band gap. 
Unlike the 1D scenario, the 2D SSH model not only features edge states but also higher-order topological corner states. 
The appearance of corner states in the 2D SSH model is due to the nontrivial product of the Zak phase along the $x-$ and $y-$ directions~\cite{Liu2019}. 
In the nontrivial unit cell of the 2D SSH thermal network, the symmetry of the point group $C_4$ ensures this condition.

The temperature profiles of the edge and corner states are depicted in Fig.~4(c). For the corner state, the temperature is localized around the four corners of the topological interfaces. In contrast, the edge state temperature distribution is along the edges of these interfaces. The positive and negative values in the temperature distribution are referenced to the ambient temperature, set at 294.15K in our analysis.

The temporal relaxation behaviors of the corner, edge, and bulk states within the 2D SSH thermal network are illustrated in Fig.~4(d).  
To analyze these behaviors, we selected specific monitoring sites: for the corner state, the monitoring site is located at the corner of the topological interface; for the edge state, it is at the middle of the interface's edge; and for the bulk state, the site is at the center of the sample. 
In line with expectations, the decay rates for both corner and edge states exhibit a constant trend over time, owing to their isolation from bulk states.

\subsection{3D SSH thermal network}
\begin{figure}[b]
	\centering
	\includegraphics[width=1\linewidth]{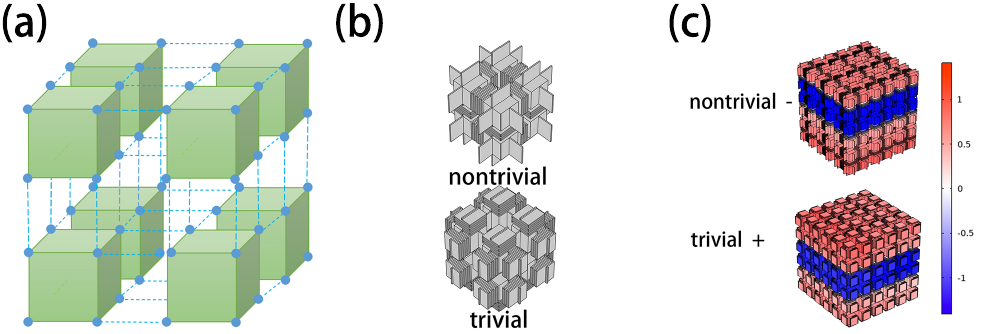}

\caption{(a) Schematic of thermal network mimicking 3D SSH model, where the solid and dashed lines indicate intra- and inter-cells thermal conductance. 
(b) Unit cells of 3D SSH thermal network for topological nontrivial and trivial phases. 
(c) Temperature profiles of 3D SSH thermal network within the first band for the nontrivial and trivial unit-cell under periodic boundary condition.}
\end{figure}

As discussed in Ref.~[\onlinecite{Liu2023}], the 3D SSH model exhibits rich topological phases depending on the magnitude relationship between $D_i$ and $D_i^\prime$ for $i=x,y,z$. 
The 3D SSH model distinguishes itself from the 2D version with two notable characteristics:
(1) The 3D SSH model reveals a layered structure of topological states, where corner states coexist with surface and hinge states. This hierarchy illustrates the complexity of the model and the intricate nature of its topological features. 
(2) Unlike the highest-order topological states (corner states), the 3D SSH model also hosts an intermediate-order topological phase. 
In this phase, the surface and hinge states are present, but the corner states are absent. A significant aspect of these intermediate-order topological states is that they are protected by band topology only within a specific part of the Brillouin zone. This selective protection makes them especially robust against perturbations along certain directions. 
The schematic of the 3D SSH model is depicted in Fig.~5(a).

Assuming periodic boundary condition, in terms of Bloch-wave solution, the tight-binding matrix $\mathbf{D}(\mathbf{k})$ of the 3D SSH thermal network can be written as 
\begin{widetext}
	\begin{equation}
		\tiny{
			\begin{aligned}
				\begin{array}{lc}
					
					\begin{pmatrix}
						-\sum_i(D_i+D^\prime_i) & {{D}_x}+{{D}^\prime_x}{{e}^{i{{k}_{x}}}} & 0 & {{D}_{x}}+{{D}^\prime_{y}}{{e}^{i{{k}_{y}}}} & {{D}_{z}}+{{D}^\prime_{z}}{{e}^{-i{{k}_{z}}}} & 0 & 0 & 0  \\
						{{D}_{x}}+{{D}^\prime_{x}}{{e}^{-i{{k}_{x}}}} & \sum_i({{D}_{i}}+{{D}^\prime_{i}}) & {{D}_{y}}+{{D}^\prime_{y}}{{e}^{i{{k}_{y}}}} & 0 & 0 & {{D}_{z}}+{{D}^\prime_{z}}{{e}^{-i{{k}_{z}}}} & 0 & 0  \\
						0 & {{D}_{y}}+{{D}^\prime_{y}}{{e}^{-i{{k}_{y}}}} & \sum_i({{D}_{i}}+{{D}^\prime_{i}}) & {{D}_{x}}+{{D}^\prime_{x}}{{e}^{-i{{k}_{x}}}} & 0 & 0 & {{D}_{z}}+{{D}^\prime_{z}}{{e}^{-i{{k}_{z}}}} & 0  \\
						{{D}_{y}}+{{D}_{y}}{{e}^{-i{{k}_{y}}}} & 0 & -{{D}_{x}}+{{D}_{x}}{{e}^{i{{k}_{x}}}} & \sum_i({{D}_{i}}+{{D}^\prime_{i}}) & 0 & 0 & 0 & {{D}_{z}}+{{D}^\prime_{z}}{{e}^{-i{{k}_{z}}}}  \\
						{{D}_{z}}+{{D}_{z}}{{e}^{i{{k}_{z}}}} & 0 & 0 & 0 & \sum_i({{D}_{i}}+{{D}^\prime_{i}}) & {{D}_{x}}+{{D}^\prime_{x}}{{e}^{i{{k}_{x}}}} & 0 & {{D}_{y}}+{{D}^\prime_{y}}{{e}^{i{{k}_{y}}}}  \\
						0 & {{D}_{z}}+{{D}^\prime_{z}}{{e}^{-i{{k}_{z}}}} & 0 & 0 & {{D}_{x}}+{{D}^\prime_{x}}{{e}^{-i{{k}_{x}}}} & \sum_i({{D}_{i}}+{{D}^\prime_{i}}) & {{D}_{y}}+{{D}_{y}}{{e}^{-i{{k}_{y}}}} & 0  \\
						0 & 0 & {{D}_{1}}+{{D}_{2}}{{e}^{i{{k}_{z}}}} & 0 & 0 & {{D}_{1}}+{{D}_{2}}{{e}^{i{{k}_{y}}}} & \sum_i({{D}_{i}}+{{D}_{i}}) & {{D}_{x}}+{{D}^\prime_{x}}{{e}^{i{{k}_{x}}}}  \\
						0 & 0 & 0 & {{D}_{z}}+{{D}^\prime_{z}}{{e}^{i{{k}_{z}}}} & {{D}_{y}}+{{D}^\prime_{y}}{{e}^{-i{{k}_{y}}}} & 0 & {{D}_{x}}+{{D}^\prime_{x}}{{e}^{-i{{k}_{x}}}} & -\sum_i({{D}_{i}}+{{D}_{i}})  
					\end{pmatrix} 
					\label{Eq:7}
				\end{array}
		\end{aligned}},
	\end{equation}
\end{widetext}
where $D_i$ is the intra-site thermal conductivity, and $D^\prime_i$ is the inter-site thermal conductivity with $i=x,y,z$. There are $2^3$ sites in a unit cell of the 3D thermal SSH network. The 3D SSH model has rich topological phases due to the additional second and third dimensions. For example, the highest-order topological phase accompanied by corner states requires $D^\prime_i>D_i$ for all $i=x,y,z$, and the intermediate-order topological phase accompanied by hinge and surface states requires $D^\prime_i>D_i$ in either two directions. We discuss the highest- and intermediate-order topological phases in thermal networks as follows. 
\begin{figure}[b]
	\centering
	\includegraphics[width=1\linewidth]{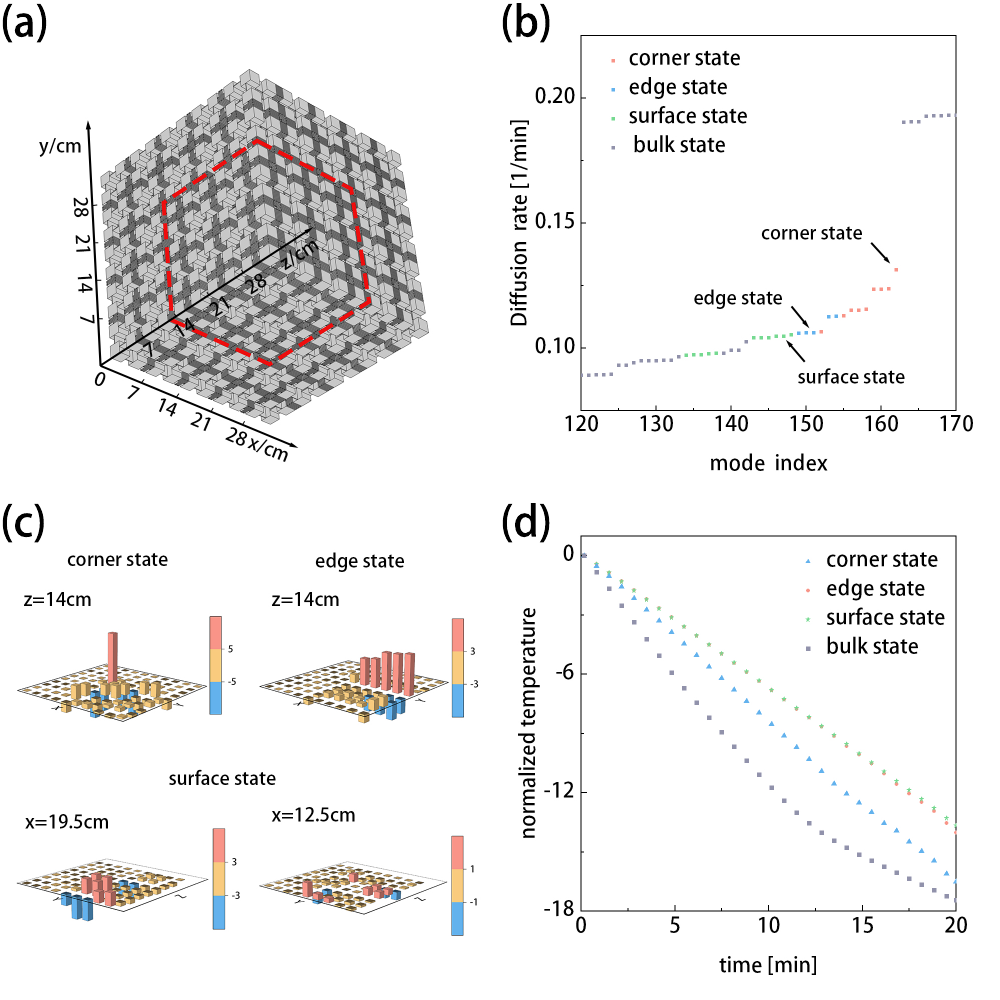}

\caption{(a) thermal network structure made up of the 3D SSH models with trivial and nontrivial unit-cells, where the dashed lines indicate the positions of topological interfaces. 
(b) Energy band spectrum of the thermal network in (a), where topological surface, hinge, and corner states appear in the band gap.
(c) Temperature profiles of the topological corner, edge, and surface states.
(d) Time evolution of bulk, surface, hinge, and corner states in (b).}
\end{figure}

\subsubsection{thermal highest-order topological phase}

In our exploration of the 3D SSH thermal network, we first focus on realizing the highest-order topological phase, characterized by the presence of corner states in a 3D system. 
To enter this phase, unit cells must exhibit greater intercell thermal conductivity compared to intracell conductance in all spatial directions. 
Figure 5(b) depicts the unit cell structure designed for the highest-order topological phase.

Following the approach used in the 1D and 2D scenarios, the topological classification of the two types of unit cells can be determined by examining the parity of the temperature profile within the lowest band, as shown in Fig.~5(c). Given that the thermal conductance is symmetric across all three directions in our setup, we anticipate the emergence of the highest-order topological phase in this thermal network.

Upon determining the topological properties of the unit cells, we proceed to assemble a finite 3D sample comprising two topologically distinct domains. 
Figure 6(a) illustrates this finite 3D thermal network, with the dashed cubic indicating the topological interface. 
The characteristic decay rate spectrum of this finite 3D sample, including the topological interface, is shown in Figure 6(b). 
As expected, the spectrum reveals the presence of surface, hinge, and corner states within the bulk band gap. 
Figure 6(c) displays the temperature profiles of these topological surfaces, hinges, and corners. 
To more effectively depict the surface state temperature profiles, we provide plots at two specific coordinates in the third dimension. 
These plots highlight how the temperature profile is localized within a surface and diminishes along the third dimension. 
Furthermore, Figure 6 (d) shows the time evolution of the bulk, surface, hinge and corner states within the 3D SSH thermal network. 
In particular, the decay rates of these topological states exhibit a more prolonged linear trend compared to the 1D and 2D cases.  

\subsubsection{thermal intermediate-order topological phase}

\begin{figure}[b]
	\centering
	\includegraphics[width=1\linewidth]{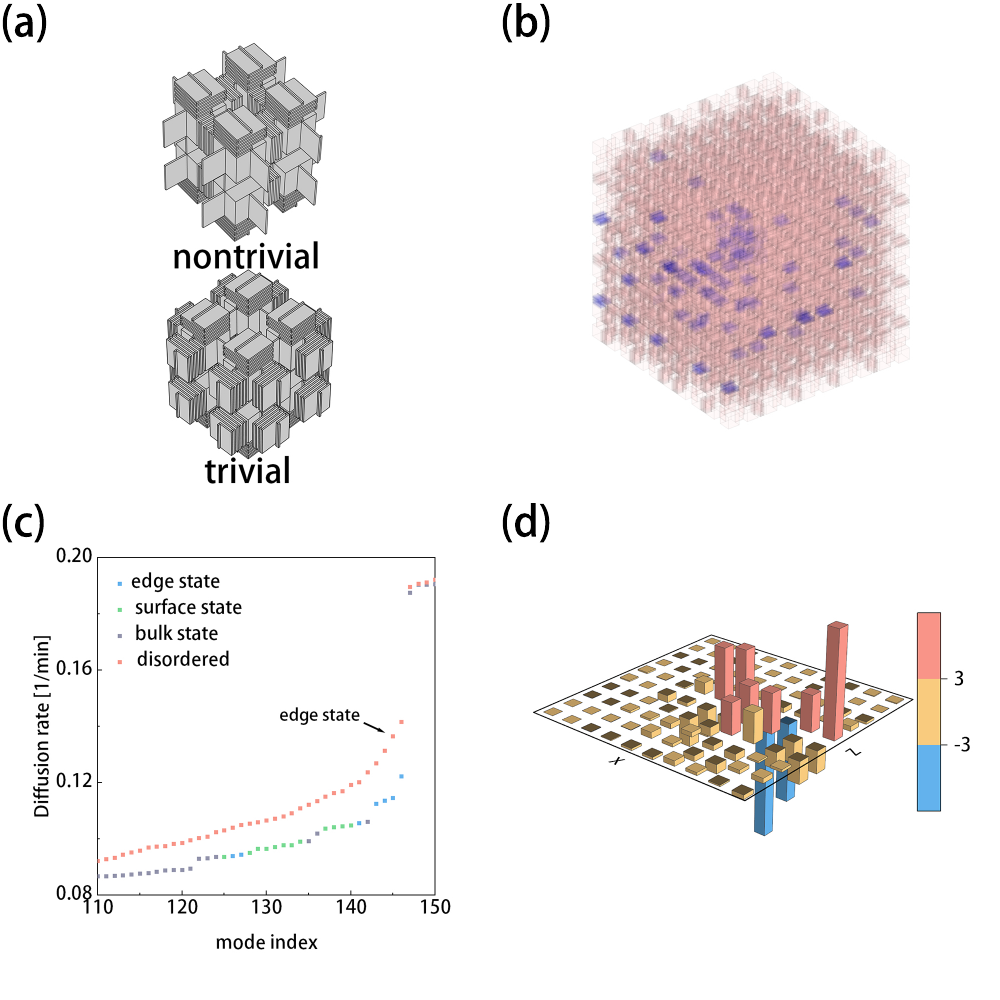}
\caption{(a) Nontrivial and trivial unit-cells for the 3D SSH thermal network in the intermediate-order topological phase. 
(b) Schematic of the perturbations on the thermal conductivity for the 3D SSH thermal network in the intermediate-order topological phase. The dark (blue) colors represent the linkage with a perturbed conductance. 
(c) Energy spectra for the 3D SSH thermal network with and without perturbations. 
The red dots are the energy spectrum in the perturbed sample.
(d) Temperature profiles of the topological edge state in the perturbed structure.}
\end{figure}

In the 3D SSH thermal network, apart from the highest-order topological phase, there exists also an intermediate-order topological phase. 
To realize this phase, condition $D^\prime_i>D_i$ must be met in two of the three spatial directions. 
As displayed in Fig.~7(a), we choose the nontrivial directions as $x$ and $y$. 
By constructing a topological interface using the unit cells from Fig.~7(a) and resolving the corresponding eigenproblem, we can determine the characteristic decay rate spectrum of the 3D SSH thermal network, as depicted in Fig.~7(b). 
Notably, in this intermediate-order phase, the corner states are absent.

An intriguing aspect of intermediate-order topological states is their robustness against perturbations in specific directions.
In our case, this robustness is against perturbations along the $z$-direction. 
Figure 7(c) presents the temperature profile of the edge states in a sample with random perturbations of up to 500\% in thermal conductivity along the $z$-directions. 
This amplitude of perturbation is comparable to the gap size.
As expected, the edge states in Fig.~7(c) is well localized. 
It is noted that the perturbation along the $z$-direction can vary in type and amplitude, potentially exceeding even the size of the band gap. 
This resilience is attributed to the fact that intermediate-order topological states are protected by the band topology within a portion of the Brillouin zone, a characteristic distinct from that of Chern insulators.  

\section{Robustness of Topological States in Thermal Networks}

\begin{figure}[b]
	\centering
	\includegraphics[width=1\linewidth]{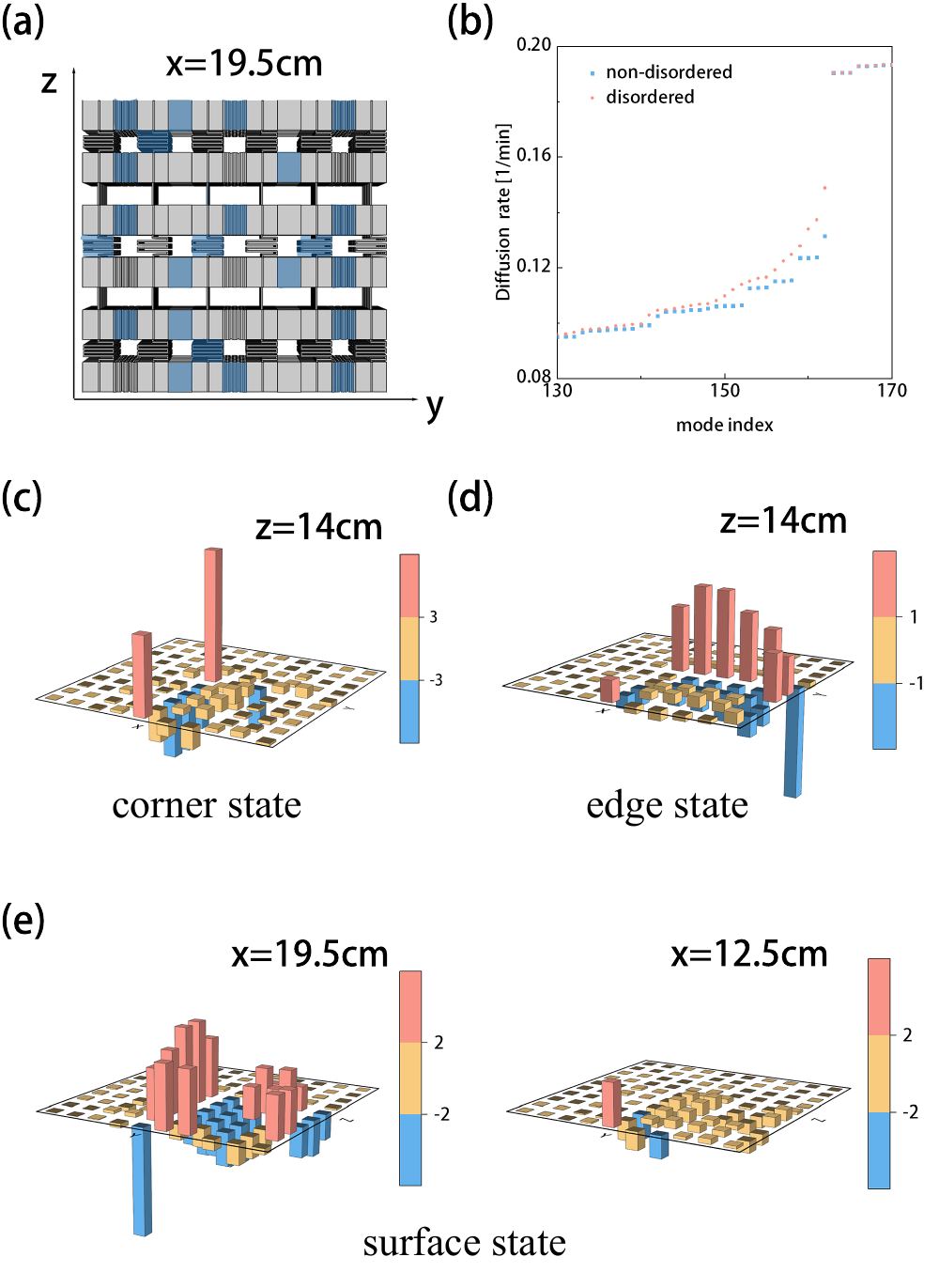}
\caption{(a) Schematic of random perturbations on the thermal conductive link in the 3D SSH thermal network in the highest-order topological phase. 
The darker color indicates the linkage with perturbations.
(b) Energy spectra of the 3D SSH thermal network for disordered and non-disordered samples.  
(c, d, e) Temperature profiles of the topological corner, hinge, and surface states in the perturbed sample of (a).}
\end{figure}

Beyond the specific direction robustness of intermediate-order topological states, the topological states in the $n$D SSH thermal network are generally protected by the magnitude of the energy gap. 
In our study, we introduce random perturbations with amplitudes of 200\% to thermal links within the 3D SSH thermal network. 
This is illustrated in Fig.~8(a).
Subsequently, we solve the eigenproblem to analyze the effects of these perturbations, with the resulting characteristic decay rate spectrum shown in Fig.~8(b). 
As observed in Fig.~8(b), despite the perturbations, the topological states remain intact. 
Figures 8 (c) to (e) show the temperature profiles of the topological corner, hinge, and surface states within the perturbed sample. Consistent with our expectations, these profiles are well localized around the topological interface, indicating the resilience of the topological states against small-scale disruptions to the thermal network.

\section{Role of Chiral Symmetry in Thermal Networks}

\begin{figure}[b]
	\centering
	\includegraphics[width=1\linewidth]{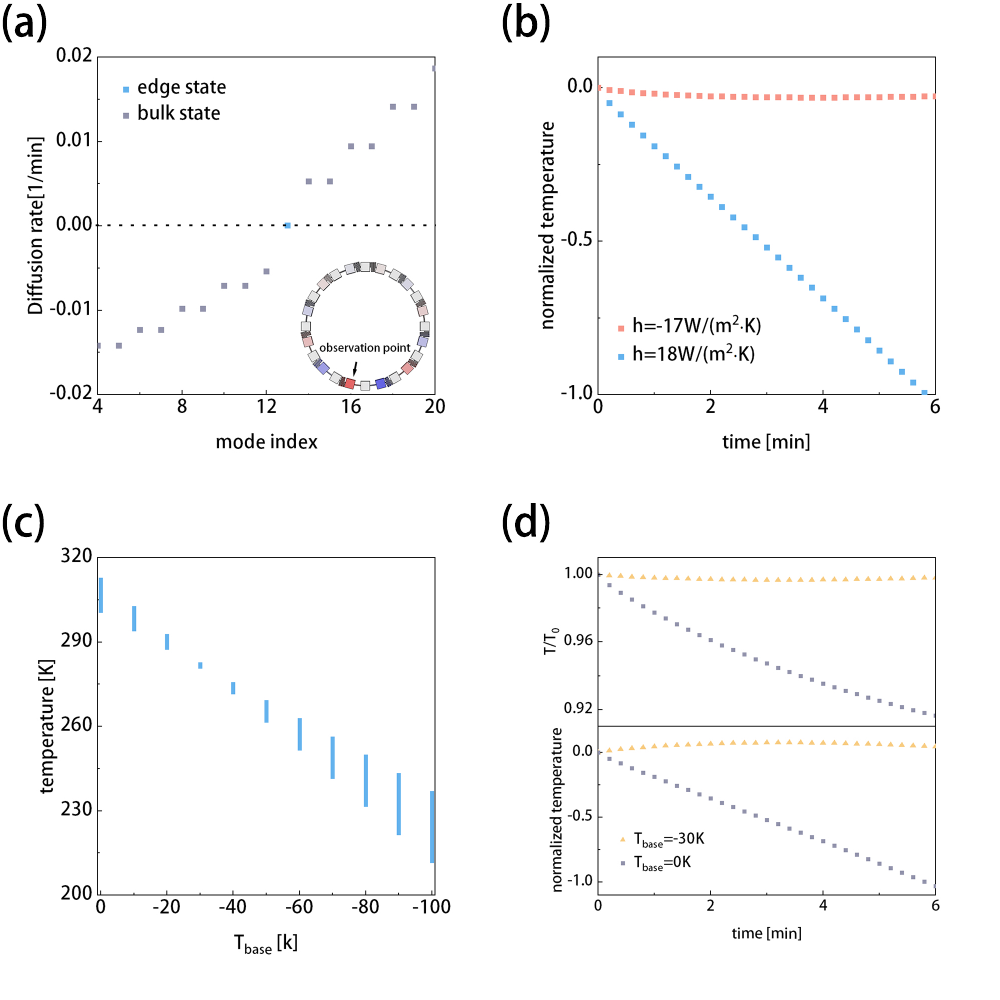}
\caption{(a) Energy spectrum of 1D SSH thermal network with negative air circulation power. Inset is the temperature profile of the zero decay rate state, where an arrow indicate the observation site for (b), (c) and (d).
(b) Comparison of the time evolution of the thermal still state with those of edge state under positive and negative air circulation power.
(c) Temperature changing range of the observation point after 6 minutes of temporal relaxation for different base temperatures of the edge state.
(d) Up panel: Time relaxation of the temperature at observation point of the edge state normalized by initial temperature for $T_\text{base}=-30$ and $0$, respectively. 
    Down panel: Time relaxation of normalized temperature at the observation point of the edge states for $T_\text{base}=-30$ and $0$, respectively. }
\end{figure}

Equation (\ref{TBM}) allows us to conceptualize the thermal network through the lens of a tight-binding model. This perspective is particularly useful in discussing the discrete symmetries that safeguard the topological states within these networks. In this paper, we specifically examine the 1D SSH thermal network, leaving a broader exploration of symmetries for future work.

A critical symmetry in the context of the 1D SSH model is chiral symmetry, characterized by the relation $\mathcal{C}^{-1}\mathcal{H}\mathcal{C}=-\mathcal{H}$. 
This symmetry ensures the anchoring of the zero-dimensional (0D) topological state at zero energy. In the thermal network, this translates to a 'still' state, defined by a zero decay rate. The presence of chiral symmetry thus not only influences the energy characteristics but also significantly impacts the dynamical behavior of the system, particularly in maintaining the stability of certain states.

To achieve a thermal still state, it is necessary to balance the off-diagonal term in Eq.~\ref{TBM}. 
One approach to achieve this balance involves utilizing air circulation. 
Assuming a periodic thermal network, the equilibrium condition can be expressed as $h=-\sum_j D^{ij}$, which results in a negative value for $h$.

For theoretical exploration, we set $h=-17\text{W}/[\text{m}^2\cdot \text{K}]$ in our simulation.
As demonstrated in Figs.~9(a) and (b), this setting allows for the realization of a thermal still state with a zero decay rate. 
The inset of Fig.~9(a) illustrates the temperature profile of this state.
To practically realize such a thermal still state, we introduce a base temperature $T_\text{base}$ at all sites of the edge state. 
For example, for the hotter sites of the edge state, we apply a negative $T_\text{base}$ to counterbalance their characteristic decay rate. 
By selecting the appropriate $T_\text{base}$, we can achieve a decay rate that is almost zero.
Figure 9(c) shows the range of temperature changes for the edge state when combined with a uniform state with temperature $T_\text{base}$ for the evolution of time in 6 minutes.
It is observed that at $T_\text{base}=-30 [\text{K}]$, the edge state exhibits minimal temperature variation, suggesting an almost still state.
In Fig.~9(d), we present the evolution of the edge state over time at $T_\text{base}=-30 [\text{K}]$.
As seen in Fig.~9(d), compared to the case with $T_\text{base}=0$, the  temperature monitored for $T_\text{base}=-30 [\text{K}]$ remains almost constant. In particular, their initial temperatures are around $280\text{[K]}$ lower than $T_\text{env}=294.15\text{[K]}$.

\section{Summary}

By analogs of $n$D SSH model using thermal networks, we demonstrate the hierarchical topological states in thermal diffusive networks up to $n=3$. 
Especially in the 3D SSH thermal network, we find an intermediate-order topological phase where the hinge and surface states exist but the corner states are absent. 
Furthermore, we show that intermediate-order topological states are robust against perturbations along a specific direction, which reflects their fractional topological nature. 
Additionally, our study delves into the role of chiral symmetry within these thermal diffusive networks. By introducing a base temperature to the thermal edge state in the 1D SSH model, we have managed to create a 'still' topological diffusive state. This state is remarkable for its near-zero decay rates, demonstrating the potential for precise thermal management. 
In general, our research lays the foundations for advanced thermal management strategies that leverage the unique properties of topological states in multiple dimensions.

\section*{Acknowledgements}
This work is supported by the Ningbo University Research Starting Funding, NSFC Grant No. 12074205, and NSFZP Grant No. LQ21A040004.

\end{document}